\documentstyle[twocolumn,prb,aps,floats]{revtex}

\newcommand{\eq}[1]{Eq.~(\ref{#1})}

\begin{document}

\draft
\twocolumn[\hsize\textwidth\columnwidth\hsize\csname %
@twocolumnfalse\endcsname

\title{Density of proper delay times in chaotic and
integrable quantum billiards}
\author{M.G.A. Crawford and P.W. Brouwer}
\address{Laboratory of Atomic and Solid State Physics, Cornell University,
Ithaca, NY, 14853}
\date{version of \today}

\widetext

\maketitle
\begin{abstract}\widetext
We calculate the density $P(\tau)$ of the eigenvalues of the Wigner-Smith
time delay matrix for two-dimensional rectangular and circular billiards
with one opening.  For long times, the density of these so-called ``proper
delay times'' decays algebraically, in contradistinction to chaotic
quantum billiards for which $P(\tau)$ exhibits a long-time cut-off.
\end{abstract}


\bigskip ]

\narrowtext

In classical mechanics, the length of time that a single particle
remains within a certain region of space is a uniquely defined
quantity. In quantum mechanics, however, the concept of ``time spent in
a region'' is not well defined, as there is no such thing as a Hermitian
``time-delay operator''.\cite{reviews} For scattering in one dimension,
Wigner has constructed a quantum-mechanical delay time in terms of the
energy-derivative of the phase shift acquired upon scattering from the
region or potential of interest.\cite{wigner} His concept was generalized
by Smith,\cite{smith60} who introduced a matrix of delay times,
\begin{equation}
Q = -i \hbar S^{\dag} \frac{\partial S}{\partial \varepsilon},
\end{equation}
where $S$ is the scattering matrix, $\varepsilon$ is the energy
of the incident particle, and $\hbar$ is Planck's constant.
In addition to being relevant for the retardation of a wave packet,
the Wigner-Smith time-delay matrix $Q$ has been shown to be related
to the capacitance,\cite{BC,gopar96} thermopower,\cite{langen98} and,
indirectly, to parametric conductance derivatives\cite{brouwer97b} and
quantum pumping.\cite{switkes,brouwer99b,zhou}

In this note, we consider the eigenvalues of the Wigner-Smith
time-delay matrix for scattering from a cavity. The eigenvalues $\tau_n$,
$1,\ldots,N$ of $Q$ are known as ``proper delay times''.\cite{smith60} A
cavity coupled to the outside world via a waveguide with $N$ propagating
channels at energy $\varepsilon$ is characterized by $N$ proper delay
times. In the semi-classical limit of large $N$, the system is described
by the density $P(\tau)$ of delay times. Our aim is to compare $P(\tau)$
for the cases where the classical dynamics of the cavity is chaotic
or integrable.  On the level of classical delay times, it is well known
that the delay time distributions are different for these two cases:
For an integrable cavity, $P_{\rm class}(\tau) \propto \tau^{-\gamma}$
has algebraic tails for large $\tau$, whereas $P_{\rm class}$ decays
exponentially for large $\tau$ if the cavity has chaotic classical
dynamics.\cite{baranger93,vicentini01} Simple arguments based on the
proximity of initial conditions to trapped periodic orbits set $\gamma
\leq 3$ for the two dimensional square and circular billiards, the two
examples of integrable cavities we consider here.

The density $P(\tau)$ of proper delay times for a chaotic cavity was
calculated by Frahm, Beenakker, and one of the authors\cite{brouwer97,extend}
using random matrix theory.\cite{beenakker97} It was found that $P(\tau)$
has a finite support,
\begin{equation}
  P(\tau) = \left\{ \begin{array}{ll}
  {1 \over 2\pi\tau^2} \vphantom{{M^M_M \over M^M_M}}
  \sqrt{(\tau_+-\tau)(\tau-\tau_-)}, & \vphantom{{M^M_M \over M^M_M}}
  \tau_- < \tau < \tau_+ \\
  0 & \mbox{otherwise},
  \end{array} \right.
  \label{eq:chaotic}
\end{equation}
where $\tau_\pm= \bar \tau(3\pm \sqrt{8})$, $\bar \tau$ being the average
delay time. The density of proper delay times for an integrable cavity
is markedly different, as we now show by consideration of the rectangular
and circular billiards in two dimensions.

\begin{figure}[h]
\begin{center}
\begin{picture}(0,0)%
\includegraphics{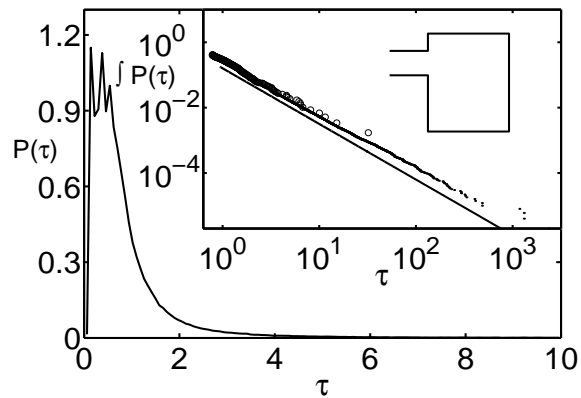}%
\end{picture}%
\setlength{\unitlength}{3947sp}%
\begingroup\makeatletter\ifx\SetFigFont\undefined%
\gdef\SetFigFont#1#2#3#4#5{%
  \reset@font\fontsize{#1}{#2pt}%
  \fontfamily{#3}\fontseries{#4}\fontshape{#5}%
  \selectfont}%
\fi\endgroup%
\begin{picture}(3600,2527)(2654,-3551)
\end{picture}
\caption{Density $P$ of proper delay times for a rectangular billiard
(shown in inset). The density $P$ is normalized to unity, $\int P(\tau)
d\tau = 1$, and the delay times $\tau$ are measured in units of their
average $\bar \tau$. The inset shows the long-time tail of the cumulative
distribution $\int_\tau P(\tau') d\tau'$, together with a (displaced)
linear fit with a slope of -1.70. The delay times were taken for an
incident wave with energy near the 17,600th level of the billiard, at
which the contact has $N=41$ propagating modes, and the density shown
was obtained after averaging over small size-preserving variations of
the shape of the billiard and the position of the lead. The circles
denote the cumulative density
for a single realization near the $3.5 \times 10^{6}$th energy level in
the billiard with $N=585$ and a fit slope of -1.66.}
\label{fig1}
\end{center}
\end{figure}

\begin{figure}
\begin{center}
\begin{picture}(0,0)%
\includegraphics{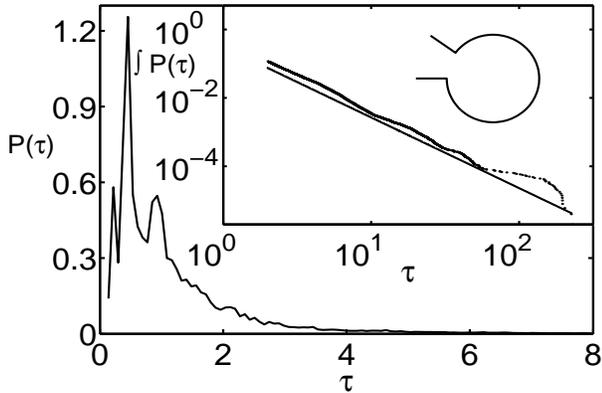}%
\end{picture}%
\setlength{\unitlength}{3947sp}%
\begingroup\makeatletter\ifx\SetFigFont\undefined%
\gdef\SetFigFont#1#2#3#4#5{%
  \reset@font\fontsize{#1}{#2pt}%
  \fontfamily{#3}\fontseries{#4}\fontshape{#5}%
  \selectfont}%
\fi\endgroup%
\begin{picture}(3924,2527)(2654,-3551)
\end{picture}
\caption{Density of proper delay times for the circle billiard (shown
in inset) and the corresponding long-time tail of the cumulative density
$\int_{\tau} P(\tau') d\tau'$. The distribution shown is obtained after
averaging over small variations of the energy around the 2250th energy
level of the billiard, for which there were $18$ conducting modes in the
point contact.}
\label{fig2}
\end{center}
\end{figure}

To find the density of proper delay times for the integrable billiards,
we solve the Schr\"odinger equation in the billiard and the leads
separately, and match the wave-functions at the billiard-lead
interface.\cite{szafer89}  Repeating this process for all possible
incident modes in the lead, a system of linear equations is formed from
which the scattering matrix $S$, its derivative $\partial S/\partial
\varepsilon$, and, hence, the time-delay matrix $Q$ can be calculated. The
billiards and the leads are shown to scale in the insets of Figs.\
\ref{fig1} and \ref{fig2}.  To improve our statistics, we have performed
an average over the position of the lead and small area-preserving
fluctuations of the billiard aspect ratio for the rectangular billiard,
and the small fluctuations of the energy $\varepsilon$ for the circular
billiard. Plots of the ensemble-averaged $P(\tau)$ for these two billiard
are shown in Figs.\ \ref{fig1} and \ref{fig2}, respectively. We have
studied the long-time asymptotics through the integrated density
$\int_{\tau} P(\tau') d\tau'$, see Figs.\ \ref{fig1} and \ref{fig2},
inset. For large $N$, the averaged density is representative of the
density $P(\tau)$ of a particular billiard, as is shown in Fig.\
\ref{fig1}, where we compare the tail of the averaged $P(\tau)$ with
the density of proper delay times for a single realization.

As can be seen from Figs.\ \ref{fig1} and \ref{fig2}, the density
$P(\tau)$ was found to decay algebraically for large $\tau$ for both
the rectangular and circular billiards.  In the rectangular billiard,
the exponent of the decay was estimated as $\gamma = 2.6$, independent of
energy in the inspected energy range $10^3 < \varepsilon < 2 \times 10^4$
(energy is measured in units such that the level spacing in the closed
billiard is one), giving a range in the number of propagating modes in
the contact ranging from $N=9$ to $N=41$.  In the circular billiard we
inspected energies in the range $10^3 < \varepsilon < 3 \times 10^3$
(corresponding to $N$ ranging from $14$ to $20$), and found that $P$
decayed with an algebraic tail with exponent $\gamma=3.05$, again
independent of energy within our accuracy. (The result that $\gamma$
is larger in the circular case could be attributed to rounding errors
in calculating Bessel functions in the large $N$ regime of interest,
introducing small random fluctuations that mimic a small randomness in the
potential, thereby suppressing long dwell times as in the chaotic regime.
The jaggedness of the small $\tau$ distribution in the circular case is
due to the restricted avenues of averaging available as compared to the
rectangular case.)  The classical distribution of delay times decays
$\propto \tau^{-\gamma}$ with $\gamma \approx 3$ in both cases.\cite{hand}
Hence, we conclude
that the density of proper delay times decays algebraically for
the integrable billiard we studied, with a power that is close, but
(for our observations) not precisely equal to the classical power
$\gamma=3$. Our results may be compared to one dimensional quasi-periodic
systems\cite{steinbach00} which also exhibit an algebraic decay of the
delay time distribution.

We wish to point out that the difference in the {\em density}
of delay times for chaotic and integrable cavities reported here
is something remarkable.  Although it is known that many quantum
properties are different for cavities with chaotic and integrable
classical dynamics, these differences usually pertain to the
statistical fluctuations, described by correlation functions, or show
up in small quantum interference corrections to a (semi)classical
background,\cite{baranger93,chang} but not in the (ensemble averaged)
densities themselves.  The only exception known to the authors is
the density of states $\rho(\varepsilon)$ in a so-called ``Andreev
quantum billiard'', an ``electron billiard'' that is connected to a
superconducting point contact. In this case, electrons that exit the
cavity and impinge on the superconductor interface are reflected as
holes and vice versa. As a result of this special reflection process,
known as Andreev reflection,\cite{andreev} $\rho(\varepsilon)$ is singular
around the Fermi energy $\varepsilon=0$:  The density of states has a gap
for a cavity with chaotic classical dynamics, while for an integrable
cavity, $\rho(\varepsilon)$ decays algebraically as $\varepsilon
\to 0$.\cite{melsen96} In fact, this feature can be connected to the
difference for the tail of the density of proper delay times $P(\tau)$
reported here via the following heuristic argument: As shown in Ref.\
\onlinecite{melsen96}, the Andreev quantum billiard has an eigenstate
at energy $\varepsilon$ precisely if the matrix product $S(\varepsilon)
S^{\dagger}(-\varepsilon)$ has an eigenvalue $-1$. Expanding $S$ around
$\varepsilon=0$ gives
\begin{equation}
S(\varepsilon) S^{\dagger}(-\varepsilon)
\approx e^{2 i \varepsilon Q/\hbar}.
\label{eq:approx}
\end{equation}
With this approximation, the condition for eigenstates in a cavity
coupled to a superconductor simplifies to\cite{foot}
\begin{equation}
e^{2 i \varepsilon \tau_n/\hbar} = -1,
\label{eq:tauepsilon}
\end{equation}
where $\tau_n$ is a proper delay time (eigenvalue of $Q$).
\eq{eq:tauepsilon}),
viewed as a constraint on the product $\varepsilon
\tau_{n}$, connects $\rho(\varepsilon)$ at small $\varepsilon$ to
$P(\tau)$ at large $\tau$.  The fact that the density of states is
gapped for the chaotic Andreev quantum billiard can then be understood as
following from the absence of a large-time tail of $P(\tau)$, cf.\
\eq{eq:chaotic}, whereas the algebraic vanishing of $\rho(\varepsilon)$
near zero energy for the integrable Andreev quantum billiard is seen
to be related to the algebraic tail of the density of delay times
in that case.\cite{foot2} We note, however, that the approximation
\eq{eq:approx} cannot be used for a quantitative estimate of the
density of states, as it becomes unreliable for energies $\varepsilon
\tau \sim \hbar$, which is precisely where the first eigenstates are
expected to appear.

While we are not aware of a method to directly measure the density of
proper delay times for an electronic system, a direct measurement of
$P(\tau)$ would be possible using the scattering of electromagnetic waves
(microwave radiation) from a metal cavity.\cite{Stein,Kudrolli,seba01}
With a suitable choice of basis, $S$ and $Q$ may be simultaneously
diagonalized.\cite{brouwer97} If the incoming electromagnetic radiation
is in a plane wave state corresponding to one of these basis vectors
and slowly modulated in intensity, the ac modulation of the outgoing
waves will be delayed by the proper delay time $\tau_{n}$ corresponding
to the incoming wave mode.\cite{smith60} In this way, the qualitative
difference between chaotic and integrable cavities should be readily
accessible from the tail of the density of proper delay times, without
further ensemble averaging.

We thank
N.W.\ Ashcroft,
C.W.J.\ Beenakker,
A.A.\ Clerk,
H.\ Schomerus, and
X.\ Waintal
for  useful discussions. This work was supported by the NSF under grants
nos.\ DMR 0086509 and DMR 9988576, by the Sloan foundation, and by the
Natural Sciences and Engineering Research Council of Canada.


\begin{thebibliography}{10}
\bibitem{hand}
  This appears to be true generically except for rectangular billiards
  of commensurate side lengths.  In this case, $\gamma$ decreases from
  three by a small amount.
\bibitem{foot} This criterion is similar to the Bohr-Sommerfeld
  quantization rule, see Ref.\ \onlinecite{melsen96}. In the
  Bohr-Sommerfeld quantization rule, the classical delay time replaces
  the Wigner-Smith proper delay time.
\end{thebibliography}
\end{document}